\documentclass[preprint,12pt]{article}
\pdfoutput=1 
\usepackage{graphicx}
\usepackage{amsfonts}
\usepackage{amsmath}
\usepackage{url}
\usepackage{comment}
\usepackage{epstopdf}
\usepackage{hyperref} 
\usepackage{color}
\usepackage{rotating}
\usepackage{natbib}
\usepackage{footnote}
\setcitestyle{authoryear,open={(},close={)}} 

\def\A{{\mathbf A}}

\def\E{{\mathbf E}}
\def\M{{\mathbf M}}

\usepackage{hyperref}

\newcommand{\clr}{\color{black}}
\newcommand{\clb}{\color{black}}

\def\PNL{{\mathcal P}}

\usepackage{graphicx}
\usepackage{subcaption}
\usepackage{listings}  
\usepackage{xcolor} 

\begin{document}

\title{Large Language Models for Time Series: an Application for Single Stocks and Statistical Arbitrage}

\author{Sebastien Valeyre \thanks{Machina Capital,  Paris, France} \thanks{Valeyre Research,  Cannes, France}  \and Sofiane Aboura \thanks{University of Paris XIII -- Sorbonne Paris Nord,  Villetaneuse, France} }



\maketitle





\begin{abstract}

	{\clr Large Language Models (LLMs)} have been adapted for time series prediction with significant success in pattern recognition. However, the common belief is that these models are not suitable for predicting financial market returns, which are known to be almost random. We aim to challenge this misconception through a counterexample. Specifically, we utilize the Chronos model from \cite{Ansari} and test both pretrained configurations and fine-tuned supervised forecasts on the largest American single stocks using data from \cite{Ordonnez}. We construct a long/short portfolio, and the performance simulation indicates that  LLMs can in reality handle time series that are nearly indistinguishable from noise, demonstrating an ability to identify inefficiencies amidst randomness and generate alpha. Finally, we compare these results with {\clr benchmark models}, highlighting significant room for improvement in LLM performance to further enhance their predictive capabilities.

\end{abstract}


\newpage

\section{Introduction}

{\clr Large Language Models (LLMs)} gained widespread popularity when ChatGPT convinced many that machines were intelligent enough to reason like humans, even though the underlying techniques merely determine the most likely sequences of words that could respond to a prompt. {\clb Besides, with regards to the integration of generative AI in the financial industry, Citadel CEO Ken Griffin says "the tech hasn't done much for hedge funds where it matters most— beating the market"\footnote{https://blinks.bloomberg.com/news/stories/T48R8PBQ99TS}. This study challenges this view by applying for the first time a LLM to a large portfolio of U.S stocks.} Recently, the introduction of the transformer architecture by \cite{Vaswani} was a key development, as it enabled fast training on large datasets. LLMs are composed of many layers of transformers and have around a billion parameters. {\clr For instance,} the T5 (Text-To-Text Transfer Transformer) architecture was introduced by \cite{Raffel}, and further advanced the field  from T5-small size with 60 million of parameters to T6-11B with 11 billion of parameters). \cite{Amatirain} describes the catalog of LLMs models from Albert to ChatGPT and classifies T5 among others. 

{\clr The incorporation of Transformers into trading, risk, and the portfolio management industry is on the verge of transforming the financial landscape due to the oncoming technological gap. The reason is straightforward: They are reaching the state-of-the-art performance of learning from time series in many domains including Traffic, Climate, Energy, and recently, Finance. The transformer is indeed a deep learning architecture that enhances considerably the analysis and processing of huge volume of diverse and complex data that are valuable for businesses and decision-makers on a real time basis. For instance, it helps making timely decision of buying and selling securities or of optimizing portfolio with risk prediction. LLMs, such like Generative Pre-Trained Transformers, are designed for specific tasks useful in the financial industry like real time translation or sentiment analysis. Still, the most challenging, crucial and valuable task remains the time series forecasting. Time series {\clb in finance} are difficult to predict because they are modeled as random walk process based on a sequence of random observations generated by a stochastic process over time. In addition, financial time series are subject to noise, temporal dependence, non-stationarity and sometimes missing values. In an efficient market it should be nearly impossible to do prediction. In the real world, there exist thousands of financial time series since several decades, but most of them are not liquid enough to be included in a sample for prediction. For example, small capitalization shares or penny stocks could be subject to less frequent or stale quotes. Econometrics models (e.g. ARIMA) are well suited for predicting financial future outcomes because they provide a clear representation of the system, which is very useful in an industry where regulators require explanations. Unfortunately, they suffer from limitations when it comes to handling large and complex data structure. Significant advancements were made from the machine learning models (e.g. Nearest-Neighbor methods, Shrinkage methods, Tree-based methods and Boosting, Support Vector Machines classifiers, Neural Networks approaches) that proved successful in capturing complex data set. On top of that, deep learning models (e.g. Long Short-Term Memory) went even further through modelling heterogeneous and long-term data to extract patterns from complex data structure, which enhances forecasting accuracy. However, they are exposed to a sort of performance saturation because they are relatively small models with hundreds of parameters. Transformers address this limitation allowing for billions of parameters to be estimated on large scale dataset (please see Wen et al. 2024 for a survey on transformers on time series). They are reshaping the field of time series forecasting given their empirical success in the so-called “one-model-for-all-datasets” (\cite{Wu}). The vanilla Transformer is built on the encoder-decoder structure where the model architecture is based on an attention mechanism to draw global dependencies between inputs and outputs (see \cite{Vaswani}). This encoder-decoder architecture has become a promising tool for predicting complex multivariate time series. LLMs share the same architecture with transformers. LLMs for time series (e.g. BERT) are able to perform time series forecasting across a variety of datasets and are capable of incorporating missing data (\cite{Gruver}). The application of such LLMs for time series prediction is an emerging field that excels in predicting time series with clear patterns (\cite{Tang}).}

\cite{Garza} introduce TimeGPT, the first foundation model for time series, capable of generating accurate predictions for diverse datasets not seen during training. They evaluate their pre-trained model against established statistical, machine learning, and deep learning methods, demonstrating that TimeGPT zero-shot inference excels in performance. \cite{Ansari} also sought to adapt these highly efficient LLMs to time series forecasting. Their approach involved representing real numbers in different bins (using a vocabulary of 4096 tokens) and training a T5 architecture on a wide range of time series data (around 90 billion observations). They produced several pretrained models of varying sizes, ranging from tiny (11 millions of parameters) to large.  TimesFM (Time Series Foundation Model) is another pretrained time-series foundation model developed by \cite{Das}. \cite{Rasul} uses a lagged features for tokenization. 
{\clr \cite{Nie} build a transformer model for multivariate time series forecasting by using segmentation of time series into subseries-level patches that served as input tokens to the Transformer and also by using channel-independence for which each channel contains a unique time series that shares the same Transformer weights through all the series.  \cite{Ekambaram} develop a neural architecture exclusively composed of multi-layer perceptron modules for multivariate forecasting and representation learning on patched time series. This novel MLP-Mixer is also patching-based and it exploits various time-series features for multivariate forecasting.} Motivated by recent advances in Large Language Models for Natural Language Processing, \cite{Ansari,Das} designed a time-series foundation model for forecasting whose out-of-the-box zero-shot performance on a variety of public datasets comes close to the accuracy of state-of-the-art supervised forecasting models for each individual dataset. Their models are based on pretraining a patched-decoder style attention model on a large time-series corpus, and can work well across different forecasting history lengths, prediction lengths and temporal granularities.

Deep learning has become very popular among researchers in finance, both for asset pricing and systematic strategies: \cite{Brugiere} tested transformers for financial time series and showed that the algorithm cannot predict returns, but can only predict squared returns. \cite{konigstein} stated that LLMs present challenges, but also opportunities, particularly for long-term financial time series forecasting. \cite{Ordonnez} implemented transformers (with only 769 parameters) but coupled them with convolutional layers and tested them on quantitative trading strategies applied to single stocks. Their results, without accounting for trading costs, were encouraging. {\clr They} applied deep learning algorithms to residual daily returns after removing common factors using techniques such as Principal Component Analysis (PCA), Fama-French factors, or Instrumental Principal Component Analysis (IPCA), implementing the methodology from \cite{Kelly}, who used financial data to constrain the eigenvectors. This approach was also developed and justified by \cite{Valeyre}. Transformers have also been used by \cite{jiang} to identify patterns in images for {\clr financial price time series}. \cite{wood} applied deep learning techniques with only a few parameters to identify the most effective trend-following indicators enhancing and timing trend-following strategies. Transformers have also been used to extract common returns, as demonstrated by \cite{Gu}. \cite{qyrana} applied a simple Short Term Reversal factor {\clr fed by} residual returns using an autoencoder-based factor model, subsequently generating a highly profitable trading strategy {\clr (i.e. through buying losing stocks based on residual returns and shorting winner stocks)}. \cite{chen} used a deep learning technique to identify anomalies in asset pricing.  

{\clr However the question remains whether a Large Language Model optimized for time series outside the field of finance is sufficiently intelligent and sensitive to capture inefficiencies and market anomalies.} {\clb Despite the burgeoning literature, there are still no papers that apply a deep learning technique with millions of parameters to financial forecasting while this is the natural number of parameters to consider in the context of LLMs.  Studies} such as \cite{chen}, \cite{jiang}, \cite{wood}, \cite{Ordonnez}, and \cite{Brugiere} utilized models with at most a few hundred parameters so their models were not real {\clr deep learning} models. This limitation was due to their inability to pre-train models with millions of parameters using large datasets from outside the financial industry. {\clb This study fills the gap by implementing deep learning algorithms with more than 11 million parameters for forecasting financial returns. This methodological approach attempts to solve the problem to what extent LLMs are intelligent enough to predict financial returns. Therefore, the contribution of this paper is to be the first that use LLMs \clb for financial time series forecasting. }

{\clr The employed methodology is two-fold. First, we} conduct a zero-shot evaluation of the predictions from pretrained and fine-tuned supervised time series foundation LLMs Chronos by \cite{Ansari}, which were pretrained on 13 datasets that do not include single stock data or stock indices, using the datasets of the residual returns of American single stocks published by \cite{Ordonnez}. The interest in using only zero-shot evaluation is  that it provides more convincing results, as overfitting is less likely in this case. Indeed overfitting is a major concern in machine learning when applied to trading, which often makes honest results appear suspect. Another interesting aspect is demonstrating how the algorithm can adapt and display 'intelligence' in trading without being specifically trained for that purpose. We simulate a portfolio that goes long {\clr the stocks whose predicted next return is positive and short the stocks whose predicted next return is negative}. Secondly, we  compare these results with {\clr the} well-known standard Short Term Reversal (STR) or trend-following approaches, as described in \cite{Jegadeesh, Jegadeesh2}. {\clr The main finding reveals that LLMs optimized on non-financial time series are still capable of producing significant forecasts for financial time series.}

\section{The methodology of our empirical backtest}

\subsection{Chronos as the LLM model}

{\clr Our objective is to assess whether one LLM-based time series model could capture inefficiencies or market anomalies in financial markets. }

{\clr We limited our test to the Chronos model because it stands out as a best-in-class LLM for time series forecasting, delivering robust performance, particularly in zero-shot and long-horizon forecasting tasks. Its adaptability and scalability make it a valuable tool across a wide range of forecasting applications.} The model used was "amazon/chronos-t5-tiny" version of the chronos which was pretrained by \cite{Ansari} on 14 datasets (Brazilian cities temperatures, Mexico city bikes, Solar, Spanish energy and weather, Taxi, USHCN, weatherbench, wiki daily, wind farms) but not on financial time series. The model has 11 millions of parameters. {\clr Chronos was presented by \cite{konigstein}. Her thorough comparative examination includes several LLM models, such as Lag-Llama from Meta (\cite{Rasul}), PatchTST from IBM (\cite{Nie}), and TSMixer from IBM (\cite{Ekambaram}). According to \cite{konigstein}, Chronos was both the easiest to implement and the most effective among the models tested, based on internal evaluations across a variety of models: local models, task-specific models, in-domain pretrained models, and out-of-domain pretrained models.} {\clr We do not claim that it is optimal or superior to simpler models without pretraining, especially given that out-of-sample results in quantitative finance papers sometimes lack of reliability and sometimes are not confirmed once the paper is published as mentioned by \cite{McLean} and \cite{Harvey}. That is why we limited our investigation to two well established and very simple benchmarks such as both the Short Term Reversal (STR) and ARIMA, while including the machine learning model of \cite{Ordonnez} as we used their data.} We used a 'context' period of 100 days so that Chronos guess the next day, knowing only the previous 100 days. We focused only on the next day return forecast. We used 100 days as a compromise to avoid running out of memory while giving Chronos a chance to capture some patterns. Additionally, we decided to limit the study to predicting the next daily returns. {\clr While forecasting weekly or monthly returns was a possible alternative, we considered that Chronos would be more effective at identifying patterns over a shorter time horizon}.

{\clr We used both the zero-shot pretrained and fine-tuned versions of Chronos. To implement the fine-tuned version, we adjusted the 11 million weights of the 'amazon/chronos-t5-tiny' model by training it on our dataset, setting $\tau$—the maximum number of training steps and one of Chronos's key parameters—to 5, 15, or 40}. Training was conducted daily during the backtest, using the data available on each respective date and starting from the weights of the previous day. {\clr The details of the parameters used for both the pretrained and fine-tuned cases are provided in Sections \ref{zeroshot} and \ref{finetune}, with additional information and the corresponding Python code available in Appendices \ref{param_pretrain} and \ref{param_finetune}}.

\subsection{Dataset description}

{\clr We use the datasets from \cite{Ordonnez}, who restricted their analysis to the most liquid stocks to mitigate trading and friction issues. Specifically, they considered stocks whose market capitalization in the prior month exceeded 0.01\% of the total market capitalization of that month, resulting in a selection of approximately the largest 550 stocks on average. {\clb This portfolio of the most liquid U.S. stocks over such a long period appears as the most appropriate sample for a deep learning application.} The total number of stocks exceeds 550 because the composition changes daily and we have a total of  9483 different stocks in the whole period. It is very common in financial studies to test strategies on the largest 500 U.S. stocks. Indeed evaluating investment strategies or identifying market anomalies often involves testing on large-cap U.S. stocks, particularly those in the S\&P 500 index. This practice is prevalent due to the S\&P 500's representation of a substantial portion of the U.S. equity market and the availability of high-quality data:  \cite{Hou} examines 452 documented anomalies in finance. The authors construct a vast data library and apply rigorous replication tests, emphasizing the use of NYSE breakpoints and value-weighted returns to mitigate the influence of microcap stocks. Their methodology underscores the importance of focusing on larger, more liquid stocks, such as those in the S\&P 500, to ensure robustness in empirical findings.  It is worth noting that \cite{Gu}, in their effort to gather sufficient data to train their 500,000\footnote{{\clr The figure of 660,000 parameters for their simple version is our own estimation, based on the code released at \url{https://github.com/dkyol/Asset-Pricing-Model}. However, this implementation does not appear to be officially validated by \cite{Gu} and may not be fully appropriate, as it assigns different weights to each stock, whereas one would typically expect a shared parameter structure across stocks}}-parameter models, explicitly stated that, "unlike the existing literature", they did not exclude micro-cap stocks. They applied no filters, resulting in an average of over 6,200 stocks per day in their dataset.} {\clr Also as we mainly use the pretrained case of Chronos, we did not need a larger dataset for the zeros shot version or the fine tuned one.}

{\clr The three different datasets of daily residual returns (IPCA, PCA and FF) provided by \cite{Ordonnez} are all based on the same raw universe of securities from the CRSP dataset spanning January 1978 to the end of 2016, but obtained using different methods for extracting the residuals.}  \cite{Ordonnez} released their datasets of  residual returns on GitHub. We utilized their three datasets whose statistics are displayed in Tab. \ref{basicstat}, each corresponding to their standard parameters with {\clr the number of factors, $K$, equal to five}: 
\begin{itemize}
	\item IPCA factors with a rolling windows of 240 months with the details described in \cite{Kelly}.  {\clr We report the following summary statistics for daily residual returns: a mean of $4.35 \times 10^{-6}$ and a standard deviation of $0.0066$.}
	\item PCA factors  with a rolling windows of 252 days.  {\clr We report the following summary statistics for daily residual returns: a mean of $2.31 \times 10^{-6}$ and a standard deviation of $0.0059$.}
	\item  FF factors  (Fama-French market, value, size, investment and profitability factors) with a rolling windows of 60 days.  {\clr We report the following summary statistics for daily residual returns: a mean of $2.95 \times 10^{-6}$ and a standard deviation of $0.0068$.}
\end{itemize}

\begin{table}[ht]
	\centering
	\centering
	\scalebox{0.7}{
		\begin{tabular}{|c|c|c|c|c|c|}  \hline
			Dataset & Mean  & SD& Skewness & Kurtosis & Total number of residual returns   \\   \hline \rule{0pt}{4ex}
			IPCA &  $4.35 \times 10^{-6}$& $0.0066$ &$1.38$ &$583$& $4079040$ \\   \hline \rule{0pt}{4ex}
			PCA &   $2.31 \times 10^{-6}$ &$0.0059$ &$1.16$ &$594$& $4079040$\\   \hline \rule{0pt}{4ex}
			FF & $2.95 \times 10^{-6}$ & $0.0068$& $1.16$&$504$  & $4079040$      \\  \hline
	\end{tabular}}
	\caption{
		Statistics of the daily residual returns of single stocks included in the 3 datasets provided by \cite{Ordonnez} }
		\label{basicstat}
	\end{table}

$K=5$ appears to be the optimal {\clr hyper parameter} according to  \cite{Ordonnez} when applying their convolution and transformer process with a {\clr performance measure, the} gross sharpe ratio, of 3.21 for Fama-French, 3.36 for the PCA and 4.16 for the IPCA. Nevertheless the sharpe {\clr ratio} in net appears to be significant only before  2006. Using residual returns allows for a less correlated dataset, which is crucial for deep learning. {\clr It is worth noting that \cite{Gu} also used IPCA, PCA and FF as benchmarks to compare the performance of the residuals produced by their autoencoder asset pricing model.}

\subsection{Description of the different simulated strategies}
Our experiment is organized in three parts. {\clr We began by testing the zero-shot version of Chronos, then we implement the fine-tuned version, and finally compared the results against three different benchmarks. We used exactly the same evaluation procedure, described in Section \ref{zeroshot}, which constructs the portfolio based on the forecasts generated by each method. Only \cite{Ordonnez}, which in one of the benchmarks, directly determined the portfolio without going through a separate forecasting step}.  

\subsubsection{Zero-shot version}
\label{zeroshot}

First, we implemented a "zero-shot evaluation", which means without any fine-tuning (or training). The weights of Chronos were not trained on financial data. Our experiment consists of, for each day, from 2001-12-26 to 2016-12-30, and for each dataset (IPCA, PCA, FF):
\begin{itemize}
	\item Computing  $\hat{\chi}_{d,i}$ , the Chronos average prediction of the next {\clr residual} daily return, derived in the Eq \ref{eq:eq3}  conditioned to a rolling window of the last 100 days. In Eq \ref{eq:eq3} we used the average of different "equiweight" scenarii $ \chi $ computed by Chronos of {\clr the forecasted residual return}, $\hat{r}_{d+1}$, knowing only $r_{d,i}...r_{d-99,i}$.  We used two possible inputs for Chronos:
	\begin{itemize}
		\item Either the last 100  residual daily returns $r_{d,i}...r_{d-99,i}$ of the single stock $i$.
		\item or the last $\hat{r}_{d,i}...\hat{r}_{d-99,i}$  the exponential moving average of the last 100  residual daily returns derived in Eq \ref{eq:eq2} with $\alpha>0$. {\clr In fact, the second possibility is merely a more general case of the first, which corresponds to the particular case where $\alpha = 0$ in Eq \ref{eq:eq2}}. We tested $\alpha$ values of 0.1, 0.2, 0.3, 0.4, 0.5, and 0.8, all different from zero. This option allows the model to account for the well-known weak negative autocorrelation of daily returns, potentially improving its forecasting ability. However, in this case, the model must outperform the Short-Term Reversal strategy described in Eq. \ref{eq:eq7} as {\clr Eq. \ref{eq:eq2} is close to Eq. \ref{eq:eq7}}. 
	\end{itemize} 
	\item Predicting of the next residual returns with $\tilde{\chi}_{d,i}$ which  is derived in Eq \ref{eq:eq4}.
	\item Calculating the weights of the portfolio $\hat{\omega}$ derived in Eq \ref{eq:eq55} which {\clr uses $\omega^\chi_d$ derived in Eq \ref{eq:eq5} which} ranks through $\Re=\text{ArgSort}$ every day $d$ the different {\clr the forecasted residual return} $\tilde{\chi}_{d,i}$  {\clr from Eq \ref{eq:eq4}}. In this method, {\clr in Eq \ref{eq:eq5}}, the median rank is withdrawn through $\frac{N}{2}$ where $N$ is the number of stocks. A normalization of the weights is derived in Eq \ref{eq:eq55} to target a gross investment of 1. This process ensures that the portfolio is 50\% long and 50\% short every day, with weights proportional to the distance in ranking from the median stock according to {\clr the forcasted residual returns} $\tilde{\chi}$ {\clr derived in Eq \ref{eq:eq4}}. \cite{Valeyre} proved that this approach is the mathematically optimal method and better than just buying the top quintile and short the bottom quintile {\clr as it is invested on a larger number of stocks with a better diversification}. A 'resized' version {\clr to manage better risk through reducing weights on volatile stocks} is also tested when the weights are also inversely proportional to the volatility as derived in Eq \ref{eq:eq555} where $\sigma$ are the standard deviation of the daily returns on the previous 100 days and $\M$ is the median.
	\item Simulating {\clr the performance of the portfolio determined by the weights in Eq \ref{eq:eq55}: We derived} $\PNL_{d+1}$, the performance of the portfolio for the next day through Eq \ref{eq:eq6}. We then reconstructed the cumulative returns  and calculated the gross Sharpe ratio, excluding any trading costs. We also simulate $\left[	\PNL_{d+1}\right]  $ for the resized version through Eq \ref{eq:eq611} {\clr which used weights derived in Eq \ref{eq:eq555}}. 
\end{itemize}

\begin{eqnarray}
	\hat{r}_{d+1,i}=\alpha \hat{r}_{d,i} + r_{d+1,i}   \label{eq:eq2}  \\
	\hat{\chi}_{d,i}= \E \left[\chi\left( \hat{r}_{d+1,i} | \hat{r}_{d,i}...\hat{r}_{d-99,i} \right)\right]\label{eq:eq3} \\
	\tilde{\chi}_{d,i}	=	\hat{\chi}_{d,i}-\alpha \hat{r}_{d,i} \label{eq:eq4} \\
	\omega^\chi_d=\Re\left[\Re\left(\tilde{\chi}_d\right)\right]-\frac{\text{N}}{2}   \label{eq:eq5}\\	
	\left[\omega^\chi_d \right]^r=\left(\Re\left[\Re\left(\tilde{\chi}_d\right)\right]-\frac{\text{N}}{2} \right) \frac{\M\left(\sigma_0...\sigma_N\right)}{\max(\sigma,\M\left(\sigma_0...\sigma_N\right)}  \label{eq:eq555}\\
	{\hat{\omega}^\chi}_{d,i}=\frac{{\omega^\chi}_{d,i}}{\sum_j{\left|{\omega^\chi}_{d,j}\right|}} \label{eq:eq55}\\
	\left[{\hat{\omega}^\chi}_{d,i}\right]^r=\frac{\left[{\omega^\chi}_{d,i}\right]^r}{\sum_j{\left|\left[{\omega^\chi}_{d,j}\right]^r\right|}} \label{eq:eq5511}\\
	\PNL_{d+1}=\sum_i{{\hat{\omega}^\chi}_{d,i} \times r_{d+1,i} }\label{eq:eq6} \\
	\left[	\PNL_{d+1}\right]=\sum_i{	\left[{\hat{\omega}^\chi}_{d,i}\right] \times r_{d+1,i} }\label{eq:eq611} 
\end{eqnarray}

.

\subsubsection{Fine-tuned version}
\label{finetune}

{\clr Secondly, we employed a very naive approach to fine-tuning from the pretrained weights, which can be highly challenging and unstable in practice \cite{Goodfellow}}. {\clr Thanks to that feed, Chronos can adapt its weights according to the properties of the financial time series}.  We trained Chronos which was initiated at the beginning of the backtest with the pretrained weights. The training was realized on a daily basis during the backtest using the {\clr available residual returns} data starting on every day with the weights of the previous day. On every day $d$, we provided as input to Chronos the updated time series available at day $d$ using the the previous 100 days. We test {\clr $5$ and $15$ as values} for $\tau$, the maximal number of steps in the iteration process. {\clr $\tau$ is a critical parameter of the algorithm, detailed in Appendix \ref{param_finetune}, governing the daily training process. If $\tau$ is set too high, Chronos may completely forget its pretrained knowledge, leading to severe overfitting—especially given the relatively small size of our dataset compared to the model’s 11 million parameters. Moreover, even a moderate value such as $\tau = 15$ is already highly computationally demanding: it required approximately one week of computation on our setup, which includes 22 Intel(R) Core(TM) i7-14700KF 3.40GHz processors, two NVIDIA GeForce RTX 4060 Ti GPUs with 80 GB of combined VRAM, and 128 GB of RAM}. We also used different values of the parameter $\alpha$ in Eq \ref{eq:eq2} {\clr which adjusts the input of Chronos}. The continuous training  led to  Chronos model to  update its weights for each day of the backtest {\clr leading to a continuous outsample result}. We then determined the portfolio weights based on the predictions using  the fine-tuned weights instead of the pretrained ones. 

We do not claim that our methodology for fine tuning is optimal to empirically determine other approaches by controling overfitting and the loss of the pretrained weights through the analysis of the statistics of the eigenvalues derived from the millions of weights (\cite{Martin19,Martin24}). However, this falls outside the scope of our current study. For instance, the drawback of our methodology is that the pretraining weights are gradually forgotten over time, which is not an ideal solution. {\clr We give more detail of the methodology we used for fine tuning in Section \ref{param_finetune} of the Appendix}.

{\clr We did perform an analysis using WeightWatcher (\cite{Martin24}) on both the pretrained and fine-tuned models, and found no signs of overfitting or underfitting. The results of this analysis (Fig \ref{fig:weighwatcher} in Sec \ref{analysisweight} of the Appendix ) which is distribution of the power law coefficient of the eigenvalues distributions layer per layer  are included in the appendix and show that the power-law exponents fall within the range of 2 to 6 in the fine-tuned case, indicating no risk of overfitting (which would be suggested by exponents below 2) or underfitting (suggested by exponents above 6). The statistics from the fine-tuned model are even more favorable than those of the pretrained model, which shows some exponents above 6, suggesting a minor risk of underfitting.}

\subsubsection{Comparison}
Third, we compare the results of Chronos to those obtained by replicating {\clr the machine learning model with a small number of 169 parameters} of \cite{Ordonnez}, {\clr who used a CNN Transformers model. Their number of paramaters is} small compared with the 11 million of the Chronos one. {\clr They used a CNN Transformers model.}  We also include  the results achieved using {\clr standard} autoARIMA from the statsforecast package ( https://pypi.org/project/statsforecast/) as it is a standard benchmark in Machine Learning, as well as the Short Term Reversal (STR) described in \cite{Jegadeesh, Jegadeesh2} which is a well-documented market anomaly that was first noted  by \cite{Fama} {\clr as the same anomaly is captured by Chronos}.

{\clr Both AutoARIMA and Short Term Reversal (STR) appear to be well-suited benchmarks, as they capture anomalies that are very similar to those detected by Chronos within our framework. \cite{Ordonnez} captures also Short Term Reversal (STR). {\clr The returns between their portfolios are correlated.} }

{\clr The Autoregressive integrated moving average (ARIMA) specification is part of the ARMA statistical models that typically predict future values based on explicit past values. The class of ARIMA models has been widely applied through \cite{Box} methodology and has proved fruitful for time series modeling according to \cite{Engle}. They are made with three components: First, an Auto-Regressive (AR) component based on a number of lagged values with order (p). Second, a Moving Average (MA) component based on the number of lagged residual terms. Third, an Integrated (q) component based a number of differencing orders used to make the data stationary on a degree (d). The auto-ARIMA specification use information criteria such like AIC or BIC to determine optimally the ARIMA parameters (p,d,q), where the best ARIMA model for the time series is one with the smallest AIC or BIC.} The AutoARIMA was also fit in a continuous way every day during the backtest using the previous 100 days $\times$ N observations and yielded to forecasts $\A_d$  and the portfolio weights ${\omega}^\A_d$ were also derived with the same methodology Eq \ref{eq:eq9}.

{\clr Short Term Reversal (STR) financial strategy has been a market anomaly that was first observed by \cite{Fama} and first documented by \cite{Jegadeesh}. It belongs to the category of price trend-following strategies in the industry of portfolio management and it captures the Short Term mean-reversion behavior. In practice this dynamic strategy requires sorting stocks into deciles according to their past-month returns in order to buy losers (stocks in the bottom decile) and sell winners (stocks in the top decile). It is a dynamic strategy because it is rebalanced on a monthly basis. It is also a robust and statistically significant strategy because \cite{Jegadeesh} unveil a monthly extra return of 2\% (see for a detailed discussion, see e.g. \cite{Da}). The Short Term Reversal (STR) strategy was }{\clr implemented using an exponential moving average of residual returns, as defined in Equation \ref{eq:eq7}, with two parameter settings: $\beta = 1 - \frac{1}{5}$ and $\beta = 1 - \frac{1}{20}$.} The portfolio of the STR is then  derived in Eq \ref{eq:eq8} with  ${\omega^\zeta }_d$, the weights of the portfolios, using the same methodology as for the Chronos forecast  where ${\text{N}}$ is the number of single stocks.

\begin{eqnarray}
	\tilde{r}_{d+1,i}=\beta \times \tilde{r}_{d,i} + r_{d+1,i}   \label{eq:eq7}  \\
	\omega^\zeta_d=\Re\left[\Re\left(-\tilde{r}_d\right)\right]-\frac{\text{N}}{2}   \label{eq:eq8}\\
	\omega^\A_d=\Re\left(\Re\left(\A_d\right)\right)-\frac{\text{N}}{2}   \label{eq:eq9} 	
\end{eqnarray}

\subsection{Performance measures}

{\clr We based our results on the gross Sharpe ratio metric of the portfolio which corresponds to the outsample annualized average of daily returns of the portfolio divided by the annualized standard deviation of daily returns, meaning that we do not consider slippage and trading costs. This metric is standard in the portfolio management literature and can be associated with the t-statistic by multiplying it by the square root of 15, corresponding to the 15 years in our sample period. If the Sharpe ratio of the portfolio is positive and the associated t-statistic exceeds 3, we can consider that the prediction methodology has succeeded in capturing some market anomalies through its forecasts. \cite{Fama15} typically assess anomalies by comparing the performance of portfolios formed on different quantiles (e.g., quintiles) of a given factor. They consider an anomaly to exist if the return spread between extreme portfolios is statistically different from zero, based on a t-statistic. In this sense, our use of the Sharpe ratio—or its associated t-statistic—is closely related to their methodology. The Sharpe ratio is particularly relevant as it provides an aggregate performance measure across the entire universe and is a simple yet widely used metric in asset management to evaluate whether a strategy is efficiently generating profits. A net Sharpe ratio of 1 is already considered excellent from the perspective of a final investor. }

\section{Empirical results}

{\clr We can see that resizing the weights of the profolio  inversely proportional to the volatility improves the Sharpe ratio for Chronos slighly, as well as for the benchmarks so that we chose to report only the results obtained from this version for all methods.}

\begin{figure}[h!]
	\centering
	\begin{subfigure}{0.45\textwidth}
		\centering
		\includegraphics[width=80mm]{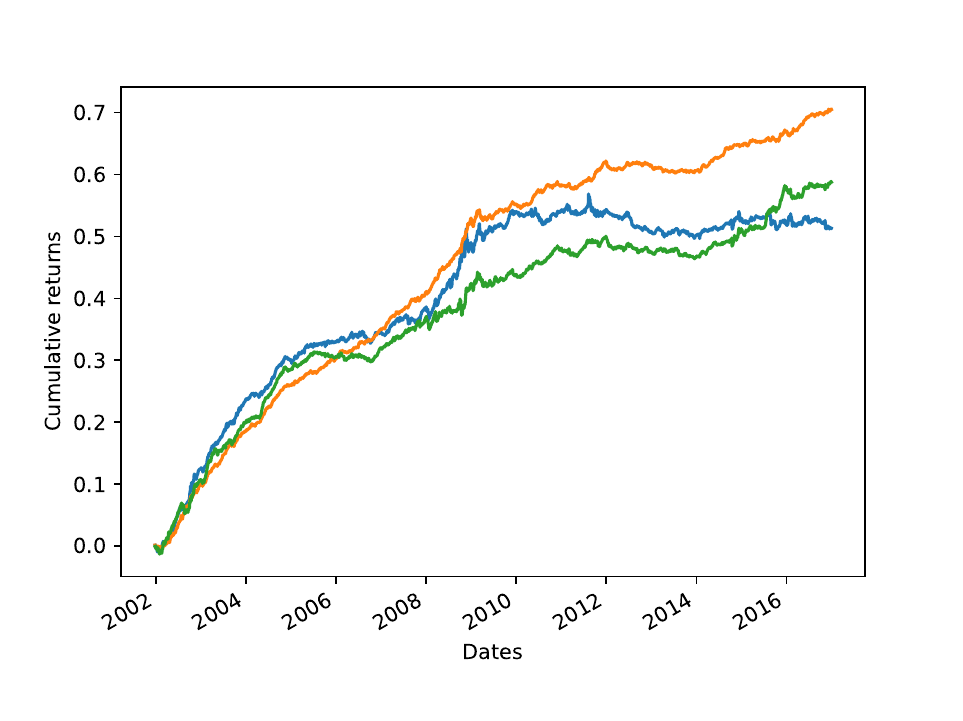}
		\caption{zero-shot pretrained prediction of Chronos. $\alpha=0.3$. The orange is for pca, the green for FF, the blue for ipca. }
		\label{fig:equity_chronos_ema_pretrained}
	\end{subfigure}
	\hfill
	\begin{subfigure}{0.45\textwidth}
		\centering
		\includegraphics[width=80mm]{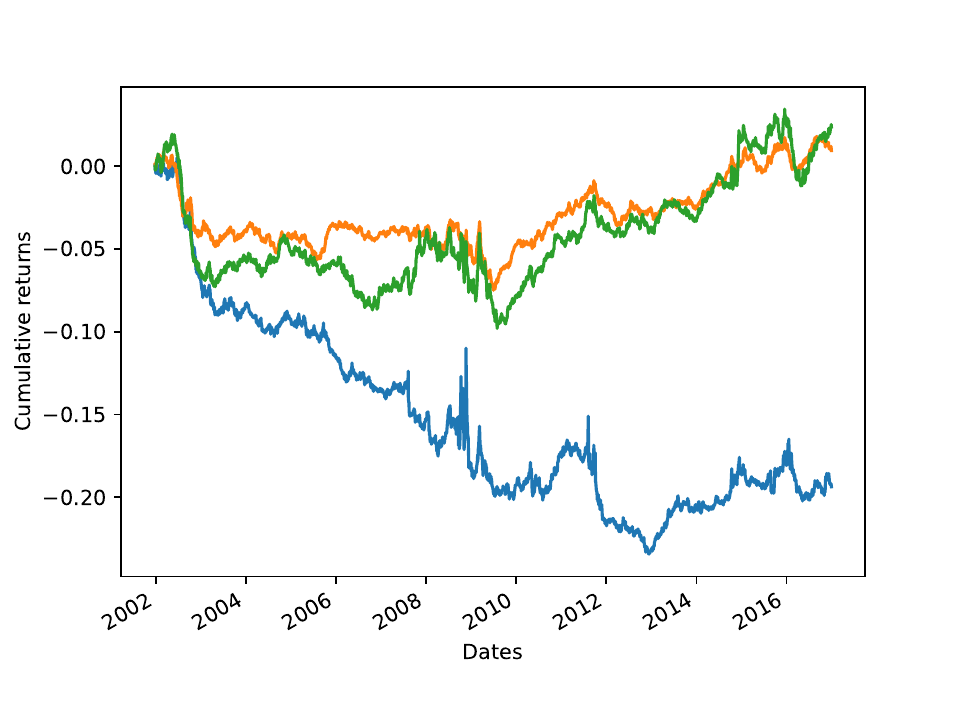}
		\caption{zero-shot pretrained prediction of Chronos. $\alpha=0$. The orange is for pca, the blue for ipca, the green for FF.}
		\label{fig:equity_chronos_withoutema_pretrained}
	\end{subfigure}
	
	\vskip\baselineskip
	\begin{subfigure}{0.45\textwidth}
		\centering
		\includegraphics[width=80mm]{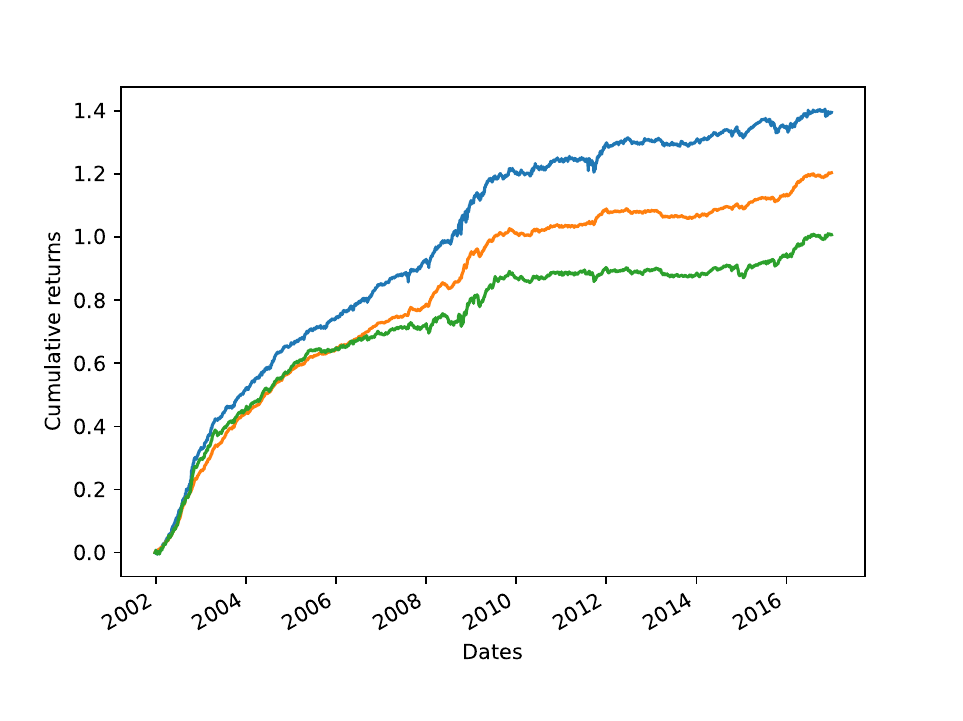}
		\caption{STR $\beta=0.8$. The blue is for ipca, the orange for pca, the green for FF.}
		\label{fig:equity_str}
	\end{subfigure}
	\hfill
	\begin{subfigure}{0.45\textwidth}
		\centering
		\includegraphics[width=80mm]{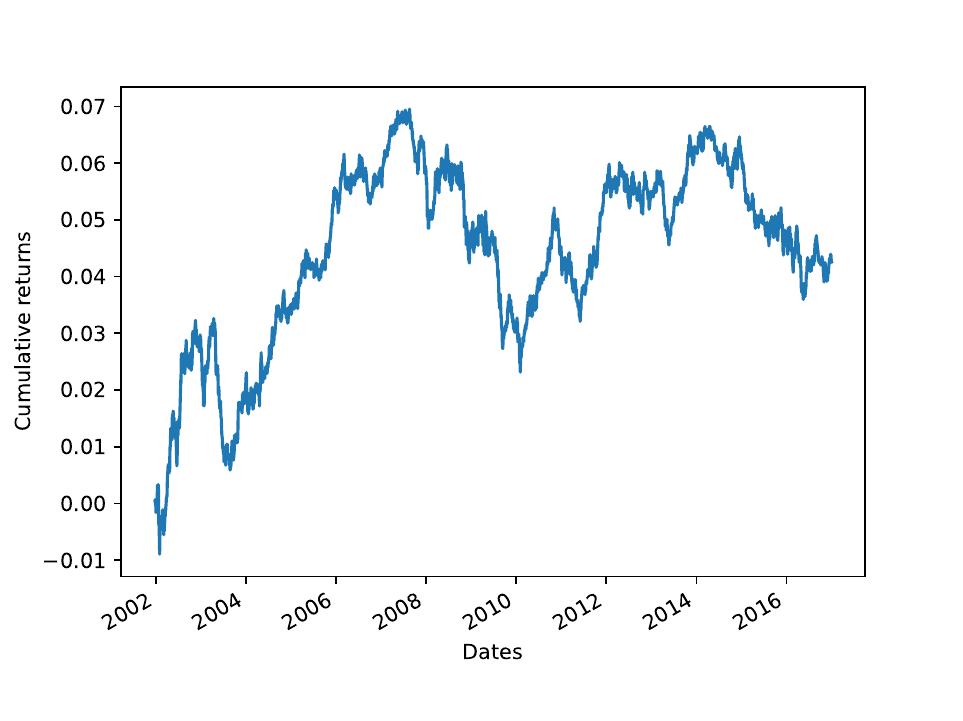}
		\caption{fined tuned Chronos. $\alpha=0$ and $\tau=15$.  the blue for pca. }
		\label{fig:equity_cnntransformers}
	\end{subfigure}
	
	\caption{Simulation of the strategy.  $\alpha$ is the parameter of the EMA in Eq \ref{eq:eq2}. $\beta$ is the parameter of the EMA in Eq \ref{eq:eq7}. $\tau$ is 'max steps' input in the fine-tuned version of Chronos. }
	\label{fig:four_graphs}
\end{figure}

We observe that the pre-trained Chronos model with {\clr the autoregressive coefficient}  $\alpha=0.3$ effectively identifies opportunities in the financial market {\clr (see Table \ref{tab:sharpe_chronos_pretrained_allapha})}, achieving a Sharpe ratio above $3.17$ for PCA over a 15-year period, which corresponds to a t-statistic of $3.17 \sqrt{15}=12.27$.  However, trading costs are prohibitive, as including a 3 basis point slippage\footnote{{\clr A transaction cost of 3 basis points per trade is consistent with the average market impact—typically between 2 and 3 basis points for very small orders—and brokerage fees of approximately 1 basis point. With $\alpha = 0.3$, the gross Sharpe ratio for PCA is 3.17, but it drops to -1.49 after accounting for a 3 basis point trading cost.} } cost per trade results in negative net Sharpe ratios. Additionally, we note a decline in profitability over time, suggesting that markets may be becoming more efficient or that opportunities are increasingly challenging to capture or that returns used to be more negatively autocorrelated before 2008 (see Figure \ref{fig:four_graphs}). However, at least until 2007, it was easier for AI to capture inefficiencies.

\begin{table}[ht]
	\centering
	\centering
	\scalebox{0.7}{
		\begin{tabular}{|c|c|c|c|c|c|c|c|}  \hline
			$\alpha$ & 0  & 0.1& 0.2 & 0.3 &0.4& 0.5& 0.8    \\   \hline \rule{0pt}{4ex}
			FF & 0.07& 1.27 &1.80 &1.84&1.39 &1.39 &-0.24  \\   \hline \rule{0pt}{4ex}
			PCA &   0.04  &2.08  &2.75 &3.17 &3.25& 2.71& 0.07 \\   \hline \rule{0pt}{4ex}
			IPCA & -0.47 & 0.68& 1.19&1.34 &1.42& 1.18& -0.81        \\  \hline
	\end{tabular}}

	\caption{
		Simulation of the  Gross sharpe ratio of the strategy {\clr with the resized version (ie. Eq \ref{eq:eq555}) to reduce risk on volatiles stocks} based on the zero-shot pretrained prediction of Chronos when using as input the exponential moving average of daily residual returns through using either the IPCA, the PCA or FF.   $\alpha$ is the parameter of the EMA from Eq \ref{eq:eq2}. The period is 2002-2016.} 
	\label{tab:sharpe_chronos_pretrained_allapha}
\end{table}

\begin{table}[ht]
	\centering
		\centering
		\scalebox{0.7}{
			\begin{tabular}{|c|c|c|c|c|c|c|}  \hline
				& FF  & PCA&IPCA    \\   \hline \rule{0pt}{4ex}
				pretrained Chronos $\alpha=0.2$ &1.80 &2.75  & 1.19 \\   \hline \rule{0pt}{4ex}
				trained Chronos $\alpha=0 $ $\tau=15$&     & 0.24  &  \\   \hline \rule{0pt}{4ex}
				trained Chronos $\alpha=0.3$ $\tau=5$&   2.12  & 3.90  &2.29  \\   \hline \rule{0pt}{4ex}
				trained Chronos $\alpha=0.3$ $\tau=15$&    &3.97 &  \\   \hline \rule{0pt}{4ex}
				resized trained Chronos $\alpha=0.3$ $\tau=15$&    &4.21 &  \\   \hline \rule{0pt}{4ex}
				trained Chronos $\alpha=0.3$ $\tau=40$&    &3.80 &  \\   \hline \rule{0pt}{4ex}
				CNN Transformer&   3.15   & 5.01  & 4.29  \\   \hline \rule{0pt}{4ex}
				STR $\beta=0.2$&   2.23  & 4.16 &  2.31 \\   \hline \rule{0pt}{4ex}
				STR $\beta=0.3$&   2.16  & 4.03 &  2.31 \\   \hline \rule{0pt}{4ex}
				resized STR $\beta=0.3$&   2.31  & 4.27 &  2.32 \\   \hline \rule{0pt}{4ex}
				STR $\beta=0.8$&   1.24  & 2.42 &  1.76 \\   \hline \rule{0pt}{4ex}
				STR $\beta=0.95$&  0.98   &1.38  &  1.20\\   \hline \rule{0pt}{4ex}
				autoARIMA &1.43   &2.10& 1.22        \\  \hline
		\end{tabular}}
	
	\caption{
		Simulation of the Gross  sharpe ratio of {\clr different strategies using {the resized version (ie. Eq \ref{eq:eq555})} to reduce risk on volatiles stocks}.  $\alpha$ is the parameter of the EMA in Eq \ref{eq:eq2}. $\beta$ is the parameter of the EMA in Eq \ref{eq:eq7}. $\tau$ is 'max steps' input in the fine-tuned version of Chronos. The period is 2002-2016.   }
	\label{tab:sharpe_chronos_pretrained_allstrat}
\end{table}

It is particularly interesting to observe that the pre-trained version with $\alpha=0$  is ineffective until 2007 but seems to work after 2008 (see Figure \ref{fig:four_graphs}). In our interpretation, Chronos is pre-trained on data where 'trend' serves as an efficient indicator, whereas in our dataset, residual returns tend to be negatively autocorrelated in the short term. We believe that $\alpha=0.3$ is optimal, as it offsets this effect, helping Chronos to overcome biases from its trend-oriented training dataset. Setting $\alpha=0.3$ artificially {\clr helps} Chronos. However, it ultimately gets very correlated to the Short-Term Reverseal (STR) strategy  with $\beta=0.3$ but fails to outperform it . In other words, when Chronos is {\clr fed} with the Exponential Moving Average (EMA), its performance does not exceed that of a zero forecast. Nevertheless, it avoids being affected by detrimental noise, which is already a positive outcome.

Table \ref{tab:sharpe_chronos_pretrained_allstrat} presents the Sharpe ratios for the fine-tuning case as well as for the benchmarks. When setting $\alpha=0$, Chronos requires fine-tuning with $\tau=15$, {\clr which is the number of iterations of the algorithm}, to achieve a Sharpe ratio of $0.24$, which corresponds to a t-statistic of $0.24 \sqrt{15} = 0.92$. It appears to work well until 2008 (see Figure \ref{fig:four_graphs}), but after that, the pre-trained configuration may have been completely forgotten due to the numerous fine-tuning processes performed since 2002. It might be interesting to test a version where there is a regular reinforcement of the pre-trained configuration to ensure it remains in memory. Additionally, the correlation with the STR strategy is not significant when $\alpha=0$ and $\tau=15$. In contrast, the CNN-Transformer appears to be primarily a linear combination of STR strategies at different time scales {\clr based on its empirical correlation with such a basktet}. This suggests that the opportunities captured by Chronos may be more complex than those driven by basic mean reversion {\clr process}. We also observe that the autoARIMA model, which is a classical benchmark in Machine Learning  underperforms  the STR model, demonstrating that fitting a model is particularly challenging when the data are nearly random and contain significant noise. This results in underperformance compared to more rigid models like STR. Finally, it is interesting to note that the optimal $\tau$ for training appears to be 15. When $\tau=40$, the Sharpe ratio decreases, suggesting that Chronos may lose some of its pre-trained intelligence.

\section{Conclusion}

Our results show that AI, specifically LLMs, can be trained on large datasets that exclude financial time series and still exhibit enough intelligence to identify opportunities in the financial market, previously considered too challenging for AI, without the risk of overfitting. Currently, AI lacks the “intelligence” to find opportunities that remain profitable when factoring in trading costs, but we can anticipate that advancements in AI may eventually make this feasible.


Nevertheless, the specialized models, which theoretically capture well-established opportunities in an optimal way, will always prove more efficient, while AI could serve as a valuable tool for identifying more complex opportunities. That {\clr assertion is supported} by the case of the strong outperformance of the STR compared to AutoARIMA, which can capture more complexity but whose noisy fit makes it suboptimal and overly erratic.



\appendix

\section{Data}

The datasets provided by \cite{Ordonnez} are  released at  \url{https://github.com/gregzanotti/dlsa-public/tree/main/residuals}

\section{Parameters of Chronos}

We downloaded the python package:

\begin{lstlisting}
!pip install git+https://github.com/amazon-science/chronos-forecasting.git
\end{lstlisting}

\subsection{Parameters of pretrained version of Chronos}
\label{param_pretrain}

We used the version "amazon/chronos-t5-tiny" with the following parameters in python:

\begin{lstlisting}
ChronosPipeline.from_pretrained( "amazon/chronos-t5-tiny",device_map="cuda", torch_dtype=torch.bfloat16)
\end{lstlisting}

\begin{lstlisting}
load_model(
model_id="google/t5-efficient-tiny",
model_type="seq2seq",
vocab_size=4096,
random_init=False,
tie_embeddings=False,
pad_token_id=0,
eos_token_id=1)
\end{lstlisting}

\begin{lstlisting}
forecast = pipeline.predict(batch_context, 1)
\end{lstlisting}

\begin{lstlisting}
predictions = np.mean(forecast.numpy(),axis=1)-alpha_chronos*np.reshape(data_train_t[-1,group*size_goup_chronos:(group+1)*size_goup_chronos],(np.shape(data_train_t[-1,group*size_goup_chronos:(group+1)*size_goup_chronos])[0],1)) #-all_timeseries[-1,:]#np.quantile(forecast.numpy(), 0.5, axis=1)
eline.predict(batch_context, 1)
\end{lstlisting}

\subsection{Parameters of Fine tuned version of Chronos}
\label{param_finetune}

{\clr Starting from the initial pretrained weights at the beginning of the backtest period, we updated the Chronos model's weights daily throughout the backtest. Each day, this update was performed by running the training procedure on 10 subgroups of the asset universe, using the previous 100 days of data. For each subgroup, training was conducted with $\tau$ iterations per day, corresponding to the "max steps" parameter in the "TrainingArguments" configuration of the "trainer.train()" method. In other words, $\tau$ successive weight updates were applied using the 'AdamW Torch Fused' gradient optimization algorithm. We tested $\tau$ values of 5, 15, and 40. The specific Python training parameters used are listed below:}
\begin{lstlisting}
chronos.ChronosConfig(
tokenizer_class='MeanScaleUniformBins',
tokenizer_kwargs={'low_limit': -15.0, 'high_limit': 15.0},
n_tokens=4096,
n_special_tokens=2,
pad_token_id=0,
eos_token_id=1,
use_eos_token=True,
model_type="seq2seq",
context_length=length_training_chronos-1,
prediction_length=1,
num_samples=20,
temperature=1,
top_k=50,
top_p=1,
)
\end{lstlisting}

\begin{lstlisting}
TrainingArguments(
output_dir=str("./output/"),
per_device_train_batch_size=32,
learning_rate=1e-3,
lr_scheduler_type="linear",
warmup_ratio=0,
optim="adamw_torch_fused",
logging_dir=str("./output/logs"),
logging_strategy="steps",
logging_steps=500,
save_strategy="steps",
save_steps=500,
report_to=["tensorboard"],
max_steps=5,#200000,
gradient_accumulation_steps=2,
dataloader_num_workers=0,#len(loaded_data),
tf32=True,  # remove this if not using Ampere GPUs (e.g., A100)
torch_compile=True,
ddp_find_unused_parameters=False,
remove_unused_columns=False,)
\end{lstlisting}

\begin{lstlisting}
shuffled_train_dataset = tch.ChronosDataset(
datasets=(tch.create_gluonts_dataset(all_timeseries,daily_dates[length_training_chronos+t:length_training_chronos+t+1])),  #list(tch.create_gluonts_dataset2(loaded_data))
probabilities=[1.0 / len(all_timeseries)] * len(all_timeseries),
tokenizer=chronos_config.create_tokenizer(),
context_length=length_training_chronos-1,
prediction_length=1,
min_past=50,
model_type="seq2seq",
imputation_method= None,
mode="training",
).shuffle(shuffle_buffer_length=100)
\end{lstlisting}

\begin{lstlisting}
trainer = Trainer(
model=model,
args=training_args,
train_dataset=shuffled_train_dataset,)
\end{lstlisting}

\begin{lstlisting}
trainer.train()
\end{lstlisting}

\subsection{Analysis by WeightWatcher}

\begin{figure}[h!]
	\centering
\includegraphics[width=80mm]{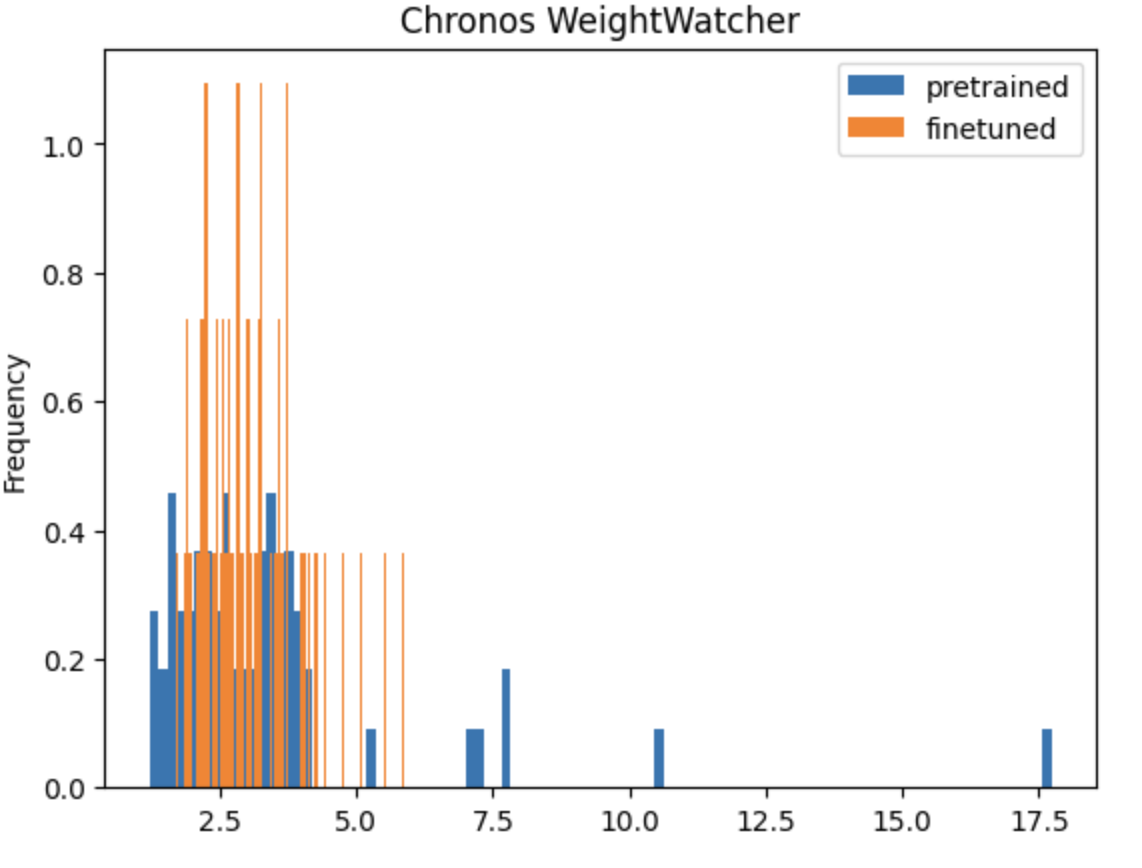}
		\caption{Analysis of the Weightwatcher (\cite{Martin24}) of the distribution of eigenvalues shows that there is no overfitting nor underfitting in both pretrained and finetuned case.	 }
		\label{fig:weighwatcher}
\end{figure}

{\clr We assessed the quality of both the pretraining and fine-tuning of Chronos using the analysis method described at \url{https://weightwatcher.ai/}. Fig \ref {fig:weighwatcher} shows that the power-law exponents fall within the range of 2 to 6 in the fine-tuned case, indicating no risk of overfitting (which would be suggested by exponents below 2) or underfitting (suggested by exponents above 6). The statistics from the fine-tuned model are even more favorable than those of the pretrained model, which shows some exponents above 6, suggesting a minor risk of underfitting.}
 
\label{analysisweight}

\section{Parameters of the CNN Transformers strategy}

We used the following major parameters provided by \cite{Ordonnez} we did not change from  \url{https://github.com/gregzanotti/dlsa-public/tree/main/config}

\begin{lstlisting}
	# Major parameters
	mode: "test"  # can be 'test' or 'estimate'
	results_tag: ""  # optional; try not to use underscores in this tag, use dashes instead
	debug: False  # set to True to turn on debug logging and file naming
	# Model parameters
	model_name: "CNNTransformer"  # name of a class defined in models folder and initialized in model folder's __init__.py
	model: {  #  contains parameter settings for __init__() function of class with name `model_name`
		lookback: 30,  # number of days of preprocessed residual time series to feed into model
		dropout: 0.25,
		filter_numbers: [1,8],
		filter_size: 2,
		attention_heads: 4,
		hidden_units_factor: 2,  # multiplicand of last item in `filter_numbers`; determines number of hidden units (e.g. 2*8 = 16)
		# hidden_units: 16,  # use either hidden_units or hidden_units_factor, but not both
		normalization_conv: True,  # normalize convolutions or not
		use_transformer: True,
		use_convolution: True,
	}
	# Data parameters
	preprocess_func: "preprocess_cumsum"  # name of a function defined in preprocess.py
	use_residual_weights: False  # use residual composition matrix to compute turnover, short proportion, etc.
	cap_proportion: 0.01  # defines asset universe: 0.01 corresponds to a residual data set
	factor_models: {  # number of factors per residual time series to test, for each factor model
		"IPCA": [5],
		"PCA": [5],
		"FamaFrench": [5],
	}
	perturbation: {  # perturbation of residual time series by noise is optional, leave empty or comment out entirely to disable
		# "noise_type" : "gaussian",
		# "noise_mean" : 0.0,
		# "noise_std_pct" : 2,
		# "noise_only" : False,
		# "per_residual" : True,
	}
	# Training parameters
	num_epochs: 100
	optimizer_name: "Adam"  # see PyTorch docs for potential optimizers
	optimizer_opts: {  # see PyTorch docs for optimizer options
		lr: 0.001
	}
	batch_size: 125
	retrain_freq: 125  # if mode=='estimate', this is the number of obs used to form a test set (chronologically after the training set)
	rolling_retrain: True  # set to False for no rolling retraining (i.e. train once, test for all data past training set)
	force_retrain: True  # force the model to be trained, even if existing weights for the model are saved on disk
	length_training: 1000  # size of rolling training window in trading days
	early_stopping: False  # employ early stopping or not
	objective: "sharpe"  # objective function: 'sharpe' or 'meanvar' or 'sqrtMeanSharpe'
	# Market frictions parameters
	market_frictions: False  # enable or disable
	trans_cost: 0  # cost in bps per txn side per equity, e.g. 0.0005
	hold_cost: 0  # cost in bps for short positions per equity per day, e.g. 0.0001
	
\end{lstlisting}


\end{document}